\documentclass[preprint2]{aastex}

\usepackage{graphicx}
\usepackage{subfigure}
\usepackage{amsmath,amssymb}
\usepackage{natbib}
\usepackage{textcomp}
\usepackage{lineno}

\newcommand{\pflux}{~cm$^{-2}$ s$^{-1}$}
\newcommand{\lsi}{LS~I~+61$^{\circ}$~303}
\newcommand{\gev}{\,GeV}
\newcommand{\tev}{\,TeV}

\shorttitle{Bright TeV flares from \lsi{}}
\shortauthors{}

\author{
S.~Archambault\altaffilmark{1},
A.~Archer\altaffilmark{2},
T.~Aune\altaffilmark{3},
A.~Barnacka\altaffilmark{4},
W.~Benbow\altaffilmark{5},
R.~Bird\altaffilmark{6},
M.~Buchovecky\altaffilmark{3},
J.~H.~Buckley\altaffilmark{2},
V.~Bugaev\altaffilmark{2},
K.~Byrum\altaffilmark{7},
J.~V~Cardenzana\altaffilmark{8},
M.~Cerruti\altaffilmark{5},
X.~Chen\altaffilmark{9,10},
L.~Ciupik\altaffilmark{11},
E.~Collins-Hughes\altaffilmark{6},
M.~P.~Connolly\altaffilmark{12},
W.~Cui\altaffilmark{13},
H.~J.~Dickinson\altaffilmark{8},
J.~Dumm\altaffilmark{14},
J.~D.~Eisch\altaffilmark{8},
A.~Falcone\altaffilmark{15},
Q.~Feng\altaffilmark{13},
J.~P.~Finley\altaffilmark{13},
H.~Fleischhack\altaffilmark{10},
A.~Flinders\altaffilmark{16},
P.~Fortin\altaffilmark{5},
L.~Fortson\altaffilmark{14},
A.~Furniss\altaffilmark{17},
G.~H.~Gillanders\altaffilmark{12},
S.~Griffin\altaffilmark{1},
J.~Grube\altaffilmark{11},
G.~Gyuk\altaffilmark{11},
M.~H\"{u}tten\altaffilmark{10},
N.~H{\aa}kansson\altaffilmark{9},
D.~Hanna\altaffilmark{1},
J.~Holder\altaffilmark{18},
T.~B.~Humensky\altaffilmark{19},
C.~A.~Johnson\altaffilmark{20},
P.~Kaaret\altaffilmark{21},
P.~Kar\altaffilmark{16},
N.~Kelley-Hoskins\altaffilmark{10},
M.~Kertzman\altaffilmark{22},
Y.~Khassen\altaffilmark{6},
D.~Kieda\altaffilmark{16},
M.~Krause\altaffilmark{10},
F.~Krennrich\altaffilmark{8},
S.~Kumar\altaffilmark{18},
M.~J.~Lang\altaffilmark{12},
G.~Maier\altaffilmark{10},
S.~McArthur\altaffilmark{13},
A.~McCann\altaffilmark{1},
K.~Meagher\altaffilmark{23},
J.~Millis\altaffilmark{24},
P.~Moriarty\altaffilmark{12},
R.~Mukherjee\altaffilmark{25},
D.~Nieto\altaffilmark{19},
S.~O'Brien\altaffilmark{6},
A.~O'Faol\'{a}in de Bhr\'{o}ithe\altaffilmark{10},
R.~A.~Ong\altaffilmark{3},
A.~N.~Otte\altaffilmark{23},
D.~Pandel\altaffilmark{26},
N.~Park\altaffilmark{27},
V.~Pelassa\altaffilmark{5},
M.~Pohl\altaffilmark{9,10},
A.~Popkow\altaffilmark{3},
E.~Pueschel\altaffilmark{6},
J.~Quinn\altaffilmark{6},
K.~Ragan\altaffilmark{1},
P.~T.~Reynolds\altaffilmark{28},
G.~T.~Richards\altaffilmark{23},
E.~Roache\altaffilmark{5},
J.~Rousselle\altaffilmark{3},
C.~Rulten\altaffilmark{14},
M.~Santander\altaffilmark{25},
G.~H.~Sembroski\altaffilmark{13},
K.~Shahinyan\altaffilmark{14},
A.~W.~Smith\altaffilmark{29},
D.~Staszak\altaffilmark{1},
I.~Telezhinsky\altaffilmark{9,10},
J.~V.~Tucci\altaffilmark{13},
J.~Tyler\altaffilmark{1},
S.~Vincent\altaffilmark{10},
S.~P.~Wakely\altaffilmark{27},
O.~M.~Weiner\altaffilmark{19},
A.~Weinstein\altaffilmark{8},
A.~Wilhelm\altaffilmark{9,10},
D.~A.~Williams\altaffilmark{20},
B.~Zitzer\altaffilmark{7}
}

\altaffiltext{1}{Physics Department, McGill University, Montreal, QC H3A 2T8, Canada}
\altaffiltext{2}{Department of Physics, Washington University, St. Louis, MO 63130, USA}
\altaffiltext{3}{Department of Physics and Astronomy, University of California, Los Angeles, CA 90095, USA}
\altaffiltext{4}{Harvard-Smithsonian Center for Astrophysics, 60 Garden Street, Cambridge, MA 02138, USA}
\altaffiltext{5}{Fred Lawrence Whipple Observatory, Harvard-Smithsonian Center for Astrophysics, Amado, AZ 85645, USA}
\altaffiltext{6}{School of Physics, University College Dublin, Belfield, Dublin 4, Ireland}
\altaffiltext{7}{Argonne National Laboratory, 9700 S. Cass Avenue, Argonne, IL 60439, USA}
\altaffiltext{8}{Department of Physics and Astronomy, Iowa State University, Ames, IA 50011, USA}
\altaffiltext{9}{Institute of Physics and Astronomy, University of Potsdam, 14476 Potsdam-Golm, Germany}
\altaffiltext{10}{DESY, Platanenallee 6, 15738 Zeuthen, Germany}
\altaffiltext{11}{Astronomy Department, Adler Planetarium and Astronomy Museum, Chicago, IL 60605, USA}
\altaffiltext{12}{School of Physics, National University of Ireland Galway, University Road, Galway, Ireland}
\altaffiltext{13}{Department of Physics and Astronomy, Purdue University, West Lafayette, IN 47907, USA}
\altaffiltext{14}{School of Physics and Astronomy, University of Minnesota, Minneapolis, MN 55455, USA}
\altaffiltext{15}{Department of Astronomy and Astrophysics, 525 Davey Lab, Pennsylvania State University, University Park, PA 16802, USA}
\altaffiltext{16}{Department of Physics and Astronomy, University of Utah, Salt Lake City, UT 84112, USA}
\altaffiltext{17}{Department of Physics, California State University - East Bay, Hayward, CA 94542, USA}
\altaffiltext{18}{Department of Physics and Astronomy and the Bartol Research Institute, University of Delaware, Newark, DE 19716, USA}
\altaffiltext{19}{Physics Department, Columbia University, New York, NY 10027, USA}
\altaffiltext{20}{Santa Cruz Institute for Particle Physics and Department of Physics, University of California, Santa Cruz, CA 95064, USA}
\altaffiltext{21}{Department of Physics and Astronomy, University of Iowa, Van Allen Hall, Iowa City, IA 52242, USA}
\altaffiltext{22}{Department of Physics and Astronomy, DePauw University, Greencastle, IN 46135-0037, USA}
\altaffiltext{23}{School of Physics and Center for Relativistic Astrophysics, Georgia Institute of Technology, 837 State Street NW, Atlanta, GA 30332-0430}
\altaffiltext{24}{Department of Physics, Anderson University, 1100 East 5th Street, Anderson, IN 46012}
\altaffiltext{25}{Department of Physics and Astronomy, Barnard College, Columbia University, NY 10027, USA}
\altaffiltext{26}{Department of Physics, Grand Valley State University, Allendale, MI 49401, USA}
\altaffiltext{27}{Enrico Fermi Institute, University of Chicago, Chicago, IL 60637, USA}
\altaffiltext{28}{Department of Applied Science, Cork Institute of Technology, Bishopstown, Cork, Ireland}
\altaffiltext{29}{University of Maryland, College Park / NASA GSFC, College Park, MD 20742, USA}

\begin{document}

\title{Exceptionally bright TeV flares from the binary \lsi{}}

\begin{abstract}
The TeV binary system \lsi{} is known for its regular, non-thermal emission pattern which traces the orbital period of the compact object in its 26.5 day orbit around its B0 Ve star companion. The system typically presents elevated TeV emission around apastron passage with flux levels between 5\,\% and 15\,\% of the steady flux from the Crab Nebula ($>300$\gev{}). In this article, VERITAS observations of \lsi{} taken in late 2014 are presented, during which bright TeV flares around apastron at flux levels peaking above $30\%$ of the Crab Nebula flux were detected. This is the brightest such activity from this source ever seen in the TeV regime. The strong outbursts have rise and fall times of less than a day. The short timescale of the flares, in conjunction with the observation of 10\tev{} photons from \lsi{} during the flares, provides constraints on the properties of the accelerator in the source.
\end{abstract}

\keywords{binaries: general --- gamma rays: general --- stars: individual (\lsi{}) --- stars: individual (VER J0240+612)--- X-rays: binaries}

\section{Introduction}

High-mass X-ray binaries (HMXBs) are a class of binary system that consist of a compact object (either a black hole or a neutron star) and a massive stellar companion and emit in X-rays. The current generation of imaging atmospheric-Cherenkov telescopes (IACTs) has facilitated the study of HMXBs which exhibit TeV emission. The class of TeV binaries is quite sparse, consisting of only a handful of sources: PSR B1259-63 \citep{2005A&A...442....1A}, LS 5039 \citep{2005Sci...309..746A}, \lsi{} \citep{Albert2006}, HESS J0632+057 \citep{2007A&A...469L...1A}, and 1FGL J1018.6-5856 \citep{2015arXiv150302711H}. TeV 2032+413 \citep{2015MNRAS.451..581L} is a new candidate TeV binary and its membership of the class is expected to be confirmed by enhanced high-energy emission during the anticipated periastron in early 2018. Of these systems, only the compact objects of PSR B1259-63 and TeV 2032+413 have been firmly identified as pulsars. There is still a large degree of ambiguity concerning the nature of the compact object within the other systems. Consequently, the fundamental mechanism responsible for the TeV emission remains uncertain. 

The orbital periods of TeV-emitting HMXBs vary from several days (LS 5039) to many years (TeV 2032+413). As the TeV emission varies strongly as a function of the orbital phase, the various sources may only have short windows during which they can be detected in the TeV regime. \lsi{} is a northern hemisphere source with a short enough orbital period to allow for regular study over multiple orbits with TeV instruments. 

Located at a distance of $\sim2$ kpc \citep{1991AJ....101.2126F}, \lsi{} is composed of a B0 Ve star and a compact object \citep{HandC1981, Casares2005}. The observed multiwavelength emission is variable at all energies and modulated with a period of $P \approx 26.5$ days, believed to be associated with the orbital motion of the binary system \citep{1982ApJ...255..210T, 1994A&A...288..519P, 1997A&A...320L..25P, Albert2006, Esposito2007, VERITASLSIDetection, Abdo2009}. Shorter timescale variability has also been detected in X-rays \citep{2009ApJ...693.1621S,2012ApJ...744..106T} and hinted at in the TeV range \citep{2013ApJ...779...88A}. Additionally, the source exhibits a periodic superorbital modulation with a period of $P_{\mbox{\scriptsize sup}} \approx 4.5$ years in H$\alpha$ emission \citep{2000A&A...358L..55Z} and in radio \citep{Gregory2002}, X-ray \citep{LiXray}, and GeV \citep{2013ApJ...773L..35A} bands. This modulation could be attributed to precessing relativistic jets if the compact object is a black hole or to cyclic variations of the B0 Ve star envelope if the compact object is a neutron star.

Radial velocity measurements show the orbit to be elliptical with eccentricity $e = 0.537\pm0.034$, with periastron occurring around phase $\phi=0.275$, apastron at $\phi=0.775$, superior conjunction at $\phi=0.081$ and inferior conjunction at $\phi=0.313$ \citep{Aragona2009}. The periastron distance between the star and the compact object is estimated to be $2.84 \times 10^{12}$\,cm (0.19\,AU) and the apastron distance to be $9.57 \times 10^{12}$\,cm (0.64\,AU) \citep{2013A&ARv..21...64D}. The inclination of the system is not precisely known but is expected to lie in the range $10^\circ$\,--~$60^\circ$ according to \citet{Casares2005}, leading to some uncertainty in the orbital parameters.

In this work, we present the results of the VERITAS campaign on \lsi{} in 2014 October\,--\,December. During this time, VERITAS observed historically bright flares from \lsi{}, with the source exhibiting flux levels a factor of 2\,--\,3 times higher than ever observed.

\section{Observations}
The VERITAS IACT array, located in southern Arizona (1.3 km a.s.l., 31$^{\circ}40^\prime$\,N, 110$^{\circ}57^\prime$\,W) consists of four 12\,m diameter Davies-Cotton design optical telescopes. VERITAS is sensitive to photons with energies from 85\gev{} to 30\tev{} and can detect a 1\,\% Crab Nebula source in approximately 25 hours \citep{parkicrc}.\footnote{\url{http://veritas.sao.arizona.edu/specifications}} For a full description of the hardware components and analysis methods utilized by VERITAS, see \citet{VERITASLSIDetection, VERITAS, KiedaVTSUpgrade}, and references therein.

\begin{figure}[ht]
\centering
\includegraphics[width=0.5\textwidth]{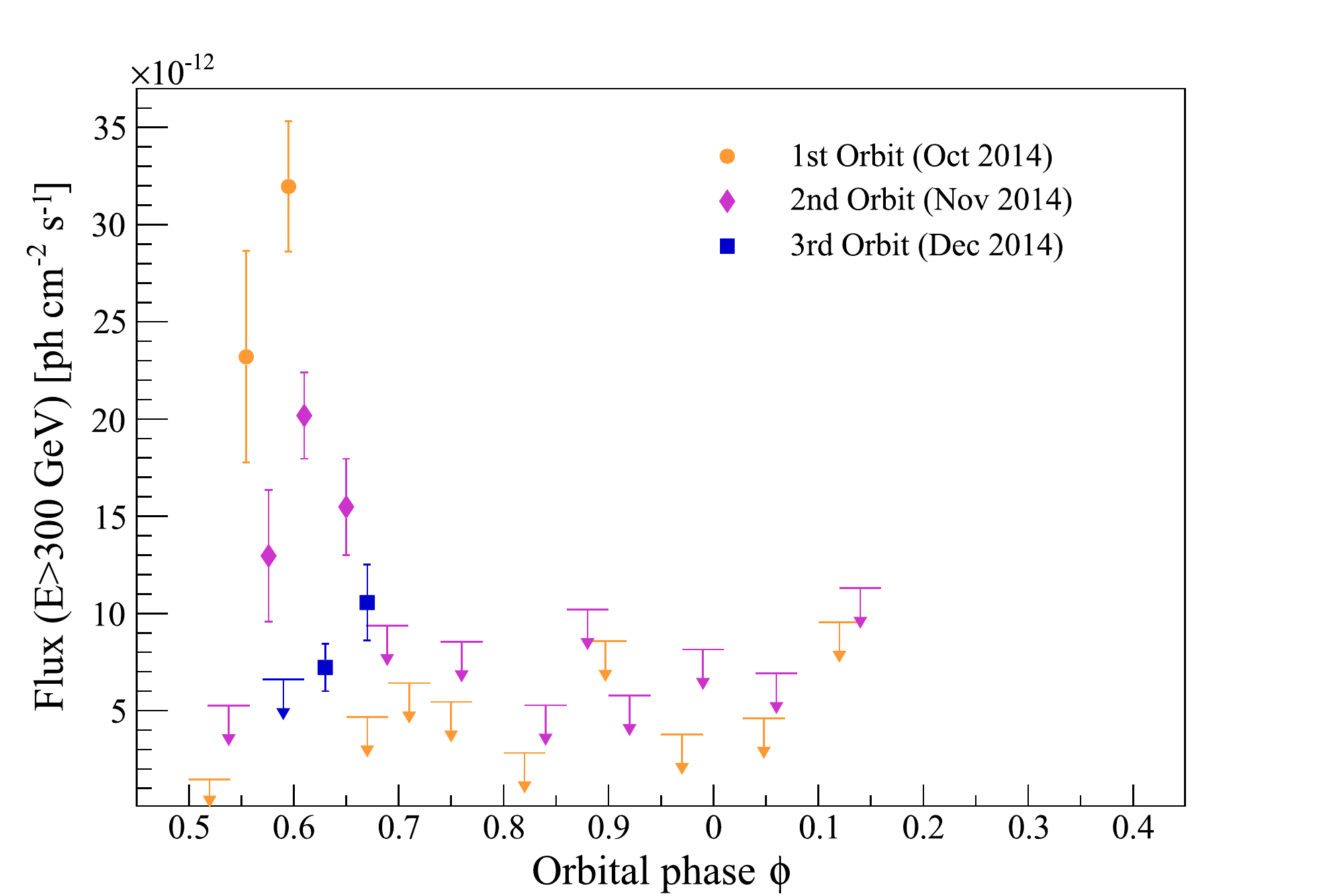}
\caption{Light curve of \lsi{} during the 2014 observation season shown as a function of orbital phase in nightly bins. The phase range is shown from 0.45 to 0.45 as the VERITAS observations commenced around a phase of $\phi=0.5$ in each orbit. The data for the first orbit (October) are shown with orange circles, while the second orbit (November) is represented by purple diamonds, and the third (December) by blue squares. Flux upper limits at the 99\% confidence level \citep[using the unbounded approach of][]{Rolke} are shown for points with significance $<3\,\sigma$ and are represented by arrows.
}
\label{f:fluxphase}
\end{figure}

In the 2014 season, VERITAS observations of \lsi{} were taken from October 16 (MJD 56946) to  December 12 (MJD 57003), comprising a total of 23.3 hours of quality-selected livetime. These observations sampled three separate orbital periods, covering the orbital phase interval $\phi = 0.5-0.2$ (see Figure~\ref{f:fluxphase} and Table~\ref{t:fluxphase}). Over the entire set of observations, a total of 443 excess events ($N_{\mbox{\scriptsize on}} = 705$, $N_{\mbox{\scriptsize off}} = 2883$, $\alpha = 0.0909$) above an energy threshold of 300\gev{} were detected above background. This is equivalent to a statistical significance of 21 standard deviations \citep[$21\sigma$, calculated using Equation 17 of][]{LiMa}. The 300\gev{} energy threshold applies to all fluxes and upper limits presented hereinafter.

\begin{deluxetable}{ccccc}
\tablecolumns{3}
\tablewidth{0pc}
\tablecaption{VERITAS observations of \lsi{} in 2014. The errors quoted on the flux are statistical only. The last column shows the pre-trials significance of the flux difference for each pair of nightly separated fluxes in units of standard deviation. The significance is only shown after the second entry of the pair. The sections show the division of the data across the three sets of observations taken in October, November, and December, respectively.}
\startdata
\hline\hline
\multicolumn{1}{c}{\vspace{-1.5ex}} & \multicolumn{1}{c}{} & \multicolumn{1}{c}{} & \multicolumn{1}{c}{} & \multicolumn{1}{c}{} \\
\multicolumn{1}{c}{Date$^1$} & \multicolumn{1}{c}{Orbital} & \multicolumn{1}{c}{Flux($>300$\gev{})} & \multicolumn{1}{c}{Duration} & \multicolumn{1}{c}{S($F_1 \neq F_2$)} \\
\multicolumn{1}{c}{[MJD]} & \multicolumn{1}{c}{phase, $\phi$} & \multicolumn{1}{c}{[$\times 10^{-11}$\pflux{}]} & \multicolumn{1}{c}{[mins]} & \multicolumn{1}{c}{[$\sigma$]} \\
\multicolumn{1}{c}{\vspace{-1.5ex}} & \multicolumn{1}{c}{} & \multicolumn{1}{c}{} & \multicolumn{1}{c}{} & \multicolumn{1}{c}{} \\ 
\hline
56946.3\phn & 0.52\phn & $<$0.15\phn\phn\phd\phn\phm{ }\phs\phm{ } & \phn24.5 & $\cdots$ \\  
56947.3\phn & 0.55\phn & \phm{$<$}2.32 $\pm$ 0.54  & \phn21.0 & 5.15 \\
56948.3\phn & 0.60\phn & \phm{$<$}3.20 $\pm$ 0.34 & \phn74.5 & 1.38 \\
56950.3\phn & 0.67\phn & $<$0.47\phn\phn\phd\phn\phm{ }\phs\phm{ } & \phn51.5 & $\cdots$ \\
56951.3\phn & 0.71\phn & $<$0.64\phn\phn\phd\phn\phm{ }\phs\phm{ } & \phn51.1 & 0.94 \\
56952.3\phn & 0.75\phn & $<$0.55\phn\phn\phd\phn\phm{ }\phs\phm{ } & \phn51.0 & 0.56 \\
56954.3\phn & 0.82\phn & $<$0.28\phn\phn\phd\phn\phm{ }\phs\phm{ } & \phn51.1 & $-$ \\
56956.3\phn & 0.90\phn & $<$0.86\phn\phn\phd\phn\phm{ }\phs\phm{ } & \phn51.5 & $-$ \\
56958.3\phn & 0.97\phn & $<$0.38\phn\phn\phd\phn\phm{ }\phs\phm{ } & \phn50.3 & $-$ \\
56960.3\phn & 0.05\phn & $<$0.46\phn\phn\phd\phn\phm{ }\phs\phm{ } & \phn50.7 & $-$ \\
56962.3\phn & 0.12\phn & $<$0.96\phn\phn\phd\phn\phm{ }\phs\phm{ } & \phn50.7 & $-$ \\ \hline
56973.2\phn & 0.54\phn & $<$0.53\phn\phn\phd\phn\phm{ }\phs\phm{ } & \phn25.5 & $\cdots$ \\
56974.2\phn & 0.58\phn & \phm{$<$}1.30 $\pm$ 0.34 & \phn34.4 & 3.63 \\
56975.2\phn & 0.61\phn & \phm{$<$}2.02 $\pm$ 0.22 & 111.3 & 1.80\\
56976.3\phn & 0.65\phn & \phm{$<$}1.55 $\pm$ 0.25 & \phn65.4 & 1.42 \\
56977.2\phn & 0.69\phn & $<$0.94\phn\phn\phd\phn\phm{ }\phs\phm{ } & \phn54.5 & 3.74 \\
56979.2\phn & 0.76\phn & $<$0.85\phn\phn\phd\phn\phm{ }\phs\phm{ } & \phn25.7 & $-$ \\
56981.3\phn & 0.84\phn & $<$0.53\phn\phn\phd\phn\phm{ }\phs\phm{ } & \phn51.3 & $\cdots$ \\
56982.3\phn & 0.88\phn & $<$1.02\phn\phn\phd\phn\phm{ }\phs\phm{ } & \phn25.7 & 0.57 \\
56983.3\phn & 0.92\phn & $<$0.58\phn\phn\phd\phn\phm{ }\phs\phm{ } & \phn33.4 & 0.62 \\
56985.2\phn & 0.99\phn & $<$0.82\phn\phn\phd\phn\phm{ }\phs\phm{ } & \phn48.8 & $-$ \\
56987.2\phn & 0.06\phn & $<$0.69\phn\phn\phd\phn\phm{ }\phs\phm{ } & \phn51.9 & $-$ \\
56989.2\phn & 0.14\phn & $<$1.13\phn\phn\phd\phn\phm{ }\phs\phm{ } & \phn51.6 & $-$ \\ \hline
57001.1\phn & 0.59\phn & $<$0.66\phn\phn\phd\phn\phm{ }\phs\phm{ } & \phn64.6 & $\cdots$ \\
57002.1\phn & 0.63\phn & \phm{$<$}0.72 $\pm$ 0.12 & 144.3 & 2.67 \\
57003.1\phn & 0.67\phn & \phm{$<$}1.06 $\pm$ 0.20 & \phn80.2 & 1.48 \\
\enddata
\label{t:fluxphase}
~\\$^1$ Start of observations
\end{deluxetable}

During the first orbit observed (in October), the source presented the largest of its flares (hereinafter ``F1''), beginning on 2014 October 17 (MJD 56947, $\phi = 0.55$). The source is not significantly detected before the onset of this flare: the flux is constrained to be less than $0.15 \times 10^{-11}$\pflux{} at a $99\%$ confidence level \citep{Rolke} on October 16 (MJD 56946, $\phi = 0.52$). The emission during F1 reached a peak flux of $(3.20 \pm 0.34) \times10^{-11}$\pflux{} on October 18 (MJD 56948, $\phi=0.60$), equivalent to approximately $30\%$ of the Crab Nebula flux above the same energy threshold of 300\gev{}. This represents the largest flux ever detected from the source. Unfortunately, observations were limited by poor weather conditions the following two nights and only recommenced on October 20 (MJD 56950), by which time the flux from the source had already decreased. During the second orbital passage in November, VERITAS detected another period of elevated flux (``F2'') from the source at similar orbital phases ($\phi = 0.55-0.65$) with peak emission of $(2.02 \pm 0.22) \times10^{-11}$\pflux{} on November 14 (MJD 56975, $\phi=0.61$).

As an initial test to show that the TeV flux is not stable, the light curves of each orbit were fitted with a constant flux model. Both F1 and F2 were found to be inconsistent with this model at the $10\,\sigma$ level using a simple $\chi^2$ test. A test for variability on a nightly timescale was then performed over the complete time range of these observations. For each pair of nightly separated fluxes ($F_1, F_2$) with statistical errors ($\sigma_1,\sigma_2$), the absolute value of the difference of the fluxes was calculated and the errors propagated using the usual variance formula. Assuming that the fluxes and errors are normally distributed, the probability that the two fluxes are not the same (i.e., $F_1 \neq F_2$) was found in terms of the standard deviation by dividing the difference by its error \citep[e.g.,][p.\,7]{dataAnalysisInHEP}:
\begin{equation}
S(F_1 \neq F_2) = \frac{|F_1 - F_2|}{\sqrt{(\sigma_1^2 + \sigma_2^2)}}.
\end{equation}
For the nights on which there was not a significant detection of the source (presented as upper limits in Figure~\ref{f:fluxphase} and Table~\ref{t:fluxphase}), the insignificant data points and their corresponding errors were used for the calculation. The most significant difference was found between the first and second nights of F1, with a pre-trials significance of $5.15\,\sigma$ corresponding to a post-trials significance of $4.66\,\sigma$ when accounting for the 12 pairs of nightly separated observations. Overall, the data are too sparsely sampled to allow a good measurement of the rise and fall times of the flares. However, the strong statistical indication of nightly variability and the sharp transition from a flux upper limit to a significantly detected flux over the course of 24 hours at the onset of F1 and F2 imply that the rise time of the flares is of the order of one day or less.

Follow-up observations conducted by VERITAS during the next month (December) covered the orbital phases of $\phi=0.59-0.67$ and detected the source at a lower flux level. The previous flares in 2014 were detected at $\phi \simeq 0.60$, but during this cycle the source reached only $(0.72 \pm 0.12) \times10^{-11}$\pflux{} at a comparable orbital phase on December 11 (MJD 57002, $\phi=0.63$). The peak emission of this cycle occurred on the following night at an orbital phase of $\phi=0.67$. The light curve of this orbit was also fitted with a constant flux model and was found to be consistent within $\sim3\,\sigma$. The observations during this month seem to exclude the type of peaked flaring behavior seen around the orbital phase $\phi \simeq 0.60$ in the previous two orbital cycles, indicating some orbit-to-orbit variations in the source.

\begin{figure}[ht]
\centering
\includegraphics[width=0.5\textwidth]{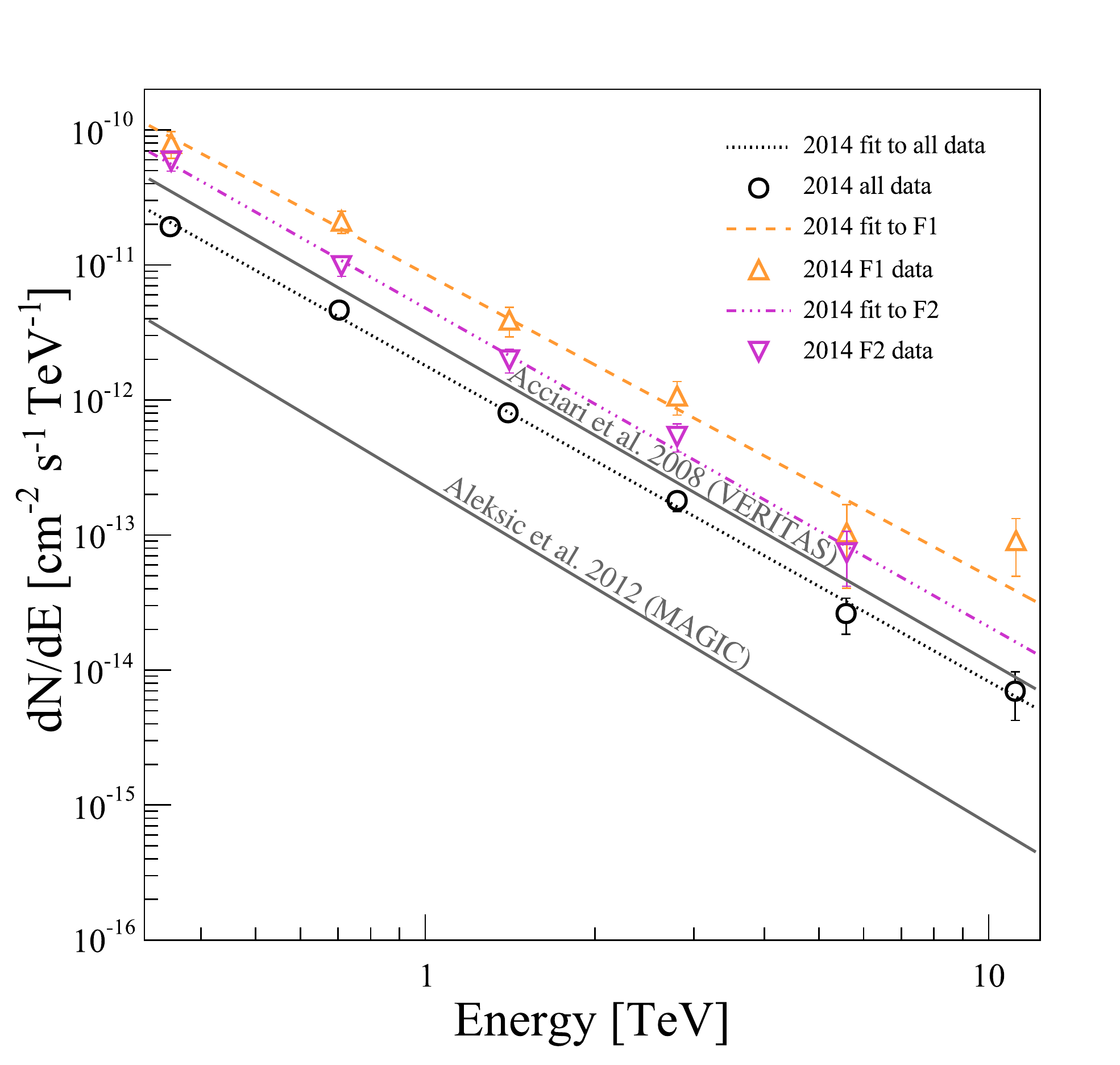}
\caption{Average and flare differential photon spectra of \lsi{} from the VERITAS 2014 observations, shown in comparison with the average spectra from \citet{VERITASLSIDetection} and \citet{Aleksic}.}
\label{spec}
\end{figure}

The average differential photon spectrum from all observations of \lsi{} during the 2014 observing season is well fit with a power law of the form
\begin{equation}
\frac{dN}{dE} = N_0 \left( \frac{E}{1\mbox{\tev{}}} \right)^{-\Gamma},
\end{equation}
in which $N_0$ is the normalization at the pivot energy of 1\tev{}, and $\Gamma$ is the spectral index. The measured parameters are consistent with past observations. Differential photon spectra were also extracted from F1 (October 17\,--\,18, MJD 56947\,--\,56948) and F2 (November 13\,--\,15, MJD 56974\,--\,56976) and show a similar spectral shape, albeit with a higher normalization constant. The parameters from the spectral fits are given in Table~\ref{t:specfits}. No spectral variability is detected within the statistical errors. All spectra are shown in Figure~\ref{spec} along with previous spectral measurements \citep{VERITASLSIDetection,Aleksic} for comparison.

An uncertainty on the energy scale of 15\,--\,25\% results in a systematic uncertainty of $\sim20\%$ on the flux normalization and $\sim40\%$ on the integral flux, assuming a spectral index of 2.34. The systematic uncertainty on the spectral index is estimated to be $\sim 0.3$, accounting for uncertainties on the collection efficiency, sky brightness, analysis cuts and simulation model. 

\begin{deluxetable}{ccc}
\tablecolumns{3}
\tablewidth{0pc}
\tablecaption{Spectral parameters of the power law fits to the observations of \lsi{} in the energy range 0.3\,--\,20\tev{}.}
\tablehead{
\colhead{Observations} & \colhead{Normalization [$\times 10^{-12}$ cm$^{-2}$ s$^{-1}$ TeV$^{-1}$]} & \colhead{Spectral index} }
\startdata
All (average) & $1.8 \pm 0.1_{\mathrm{stat}} \pm 0.4_{\mathrm{sys}}$ & $2.34 \pm 0.07_{\mathrm{stat}} \pm 0.3_{\mathrm{sys}}$ \\
F1 (Oct 17\,--\,18) & $8.6 \pm 1.0_{\mathrm{stat}} \pm 1.7_{\mathrm{sys}}$ & $2.24 \pm 0.12_{\mathrm{stat}} \pm 0.3_{\mathrm{sys}}$ \\
F2 (Nov 13\,--\,15) & $4.8 \pm 0.4_{\mathrm{stat}} \pm 1.0_{\mathrm{sys}}$ & $2.36 \pm 0.12_{\mathrm{stat}} \pm 0.3_{\mathrm{sys}}$ \\
\enddata
\label{t:specfits}
\end{deluxetable}

The highest energy gamma-ray candidates observed were detected during the peak night of F1 with energies between about $10$ and $13$\tev{}. There are no events in the background region with energies above 4\tev{} on this night, so it is assumed that the contribution of the background at $\sim10$\tev{} is negligible.

During these observations, the source was also monitored by the \emph{Fermi}-LAT (0.1\,--\,300\gev{}), the \emph{Swift}-XRT (0.2\,--\,10 keV), and both the RATAN and AMI radio instruments (4/6\,--\,15 GHz). In addition, H$\alpha$ monitoring of the system took place at the Ritter Observatory in Toledo, Ohio (USA). After F2 was detected by VERITAS, an Astronomer's Telegram\footnote{\url{www.astronomerstelegram.org}} \citep{2014VTSATEL} was released, notifying the astronomical community of the historic flux levels and triggering more intense observations by multiwavelength partners, as well as additional observations with the MAGIC TeV observatory. The results of this multiwavelength campaign are under analysis and will be presented in a future publication. 

\section{Discussion and Conclusion}
The nature of the compact object in\linebreak \lsi{} is not firmly established and, as a result, proposed emission mechanisms for the system cover a range of possibilities. These mechanisms fall into two main categories: microquasar ($\mu$Q) and pulsar binary (PB). In the $\mu$Q scenario, non-thermal particle-acceleration processes occur in the jet of an accreting compact object (see \citeauthor{2009IJMPD..18..347B} \citeyear{2009IJMPD..18..347B} for a review, or \citeauthor{Massi2001} \citeyear{Massi2001}; \citeauthor{Massi2013} \citeyear{Massi2013}; \citeauthor{2015A&A...575L...9M} \citeyear{2015A&A...575L...9M} for the specific case of \lsi{}). The pulsar binary scenario utilizes the presence of a shocked wind in which particle acceleration is the result of the interaction between the stellar and the pulsar winds \citep[e.g.,][]{MaraschiandTreves,Dubus,Dhawan2006}. While some versions of both models utilize a hadronic primary population, the  majority of both model types employ leptonic origins for the observed non-thermal emission. In a leptonic scenario, the TeV emission is the result of inverse-Compton (IC) scattering of electrons accelerated in the jet ($\mu$Q) or at the shock front (PB).

\citet{Paredes-Fortuny2014} present a general pulsar wind shock scenario with an inhomogeneous stellar wind in which the B0 Ve star disc is disrupted and fragments. The resulting clumps of the disc fall into the shock region, pushing the shock closer to the pulsar. The reduction in size of the pulsar wind termination shock could allow for increased acceleration efficiency on the timescale of a few hours, depending on the size and density of the disc fragments. Such a scenario could account for the exceptionally bright TeV flares and orbit-to-orbit variations seen in \lsi{}.

Regardless of the primary mechanism to generate the emission, \citet{2008MNRAS.383..467K} provide a prescription to calculate model-independent limits on the magnetic field strength and the efficiency of the accelerator within an IC scenario. Given the temperature $T=2.25\times10^4$\,K \citep{2013A&ARv..21...64D} of the B0 Ve star in \lsi{}, the average energy of the stellar photons is $3kT \approx 6$\,eV, and the IC scattering takes place deep in the Klein-Nishina regime, in which most of the electron energy is transferred to the scattered photons. Thus, the presence of $\sim10$\tev{} photons requires electrons with an energy of at least $10$\tev{} in the emitter, as well as forcing the acceleration time to be less than the cooling time. Two example cases will be used in the following discussion: the conservative case of a 10\tev{} electron which must be present in the emitter, and a more optimistic case of a 20\tev{} electron as there is no evidence of a cutoff in the photon spectrum. Following the calculations of \citet{2008MNRAS.383..467K}, the acceleration timescale of the electrons can be expressed as
\begin{equation}
t_{\mbox{\scriptsize{acc}}}  = \eta_{\mbox{\scriptsize{acc}}} r_L c^{-1} \approx 0.1 \eta_{\mbox{\scriptsize{acc}}} \left( \frac{E}{1\mbox{\,TeV}} \right) \left( \frac{B}{1\mbox{\,G}} \right)^{-1} \mbox{\,s},
\label{taccel}
\end{equation}
where $r_L$ is the Larmor radius of the electron, $\left( \frac{E}{1\mbox{\,\scriptsize TeV}} \right)$ is the energy of the electron in units of TeV, $\left( \frac{B}{1\mbox{\,\scriptsize G}} \right)$ is the magnetic field strength in units of Gauss, and $\eta_{\mbox{\scriptsize{acc}}} > 1$ is a parameter describing the efficiency of the accelerator (in general $\eta_{\mbox{\scriptsize{acc}}} \gg 1$). The characteristic cooling time of electrons in the Klein-Nishina regime is given by
\begin{equation}
t_{\mbox{\scriptsize{KN}}} \approx 170 \left( \frac{w_r}{100\mbox{\,erg cm$^{-3}$}} \right )^{-1} \left( \frac{E}{1\mbox{\,TeV}} \right)^{0.7} \mbox{\,s}
\end{equation}
where $w_r = L_\star / \left ( 4 \pi d^2 c \right )$ is the energy density of the starlight close to the compact object in which $L_\star$ is the luminosity of the optical star and $d^2$ is the distance between the emitter and the optical star. This distance cannot be much more than the orbital distance as otherwise the periodic component of the emission would be washed out, therefore the emitting region is assumed to be located close to the compact object. The temperature $T = 2.25 \times 10^4$\,K and radius $R = 10R_\sun$ \citep{2013A&ARv..21...64D} of the star in this system results in a total luminosity of $L_\star \approx 9 \times 10^{37}$\,erg s$^{-1}$.

The synchrotron cooling time is given by
\begin{equation}
t_{\mbox{\scriptsize{sy}}} \approx 4\times10^2 \left( \frac{B}{1\mbox{\,G}} \right)^{-2} \left( \frac{E}{1\mbox{\,TeV}} \right)^{-1} \mbox{\,s}.
\end{equation}
The relation $t_{\mbox{\scriptsize{KN}}} < t_{\mbox{\scriptsize{sy}}}$ can be set due to the fact that IC losses in the Klein-Nishina regime allow for the hard electron spectra (harder than $2$) necessary to produce hard gamma-ray spectral indices (from $2$ to $2.5$). Thus, the magnetic field in the emitter is constrained by the relation
\begin{equation}
{B < \frac{4.2 \times 10^{12}}{\sqrt{\pi}} \left( \frac{d}{\mbox{cm}} \right)^{-1} \left( \frac{E}{1\mbox{\,TeV}} \right)^{-0.85} \,\mbox{\,G}}.
\end{equation}
Using $\left( \frac{E}{1\mbox{\,\scriptsize TeV}} \right) = 10$ gives a value of $B \lesssim 0.03$\,G and using $\left( \frac{E}{1\mbox{\,\scriptsize TeV}} \right) = 20$ results in $B \lesssim 0.02$\,G at apastron (close to the position in the orbit at which the flares were detected).

A fundamental condition is that the Larmor radius of the electrons must be considerably less than the linear size of the emitter. As before, the distance between the compact object and the emitter is used as an estimate of the linear size of the emitter. Rearranging Equation~\ref{taccel} to get $r_L \approx 0.1 c \left( \frac{E}{1\mbox{\,\scriptsize TeV}} \right) \left( \frac{B}{1\mbox{\,\scriptsize G}} \right)^{-1} \mbox{\,cm}$ and taking the ratio $\frac{r_L}{d}$ gives a value of $\sim10\%$ of the total system size at apastron at the upper limit of the magnetic field strength ($B=0.03$\,G) in the case of 10\tev{} electrons. The Larmor radius of 20\tev{} electrons is already $\sim31\%$ of the system size at apastron at the upper limit of the magnetic field strength ($B=0.02$\,G). A lower magnetic field strength results in a larger Larmor radius. Given that the Larmor radius is such a large fraction of total system size, the leakage time of the electrons (not considered in this simple model) may already be the shortest timescale of the system, especially for higher-energy electrons.

The constraints are strongly dependent on the assumed location of the emitter, which has been taken to be coincident with the compact object in order to derive these limits. If the emitter is located farther from the star, the upper limit on the magnetic field is reduced even further and the orbital periodicity would be washed out. Regardless of the location of the emitter, the simple Klein-Nishina scattering assumed in this discussion implies a very hard synchrotron spectrum in the X-ray band. If the same population of electrons that is responsible for the VHE emission also produces the X-ray emission, photon indices softer than 1.3 in the X-ray regime \citep{pkaricrc} may challenge this simple scenario, which requires an electron spectral index harder than 2 to produce hard gamma-ray spectral indices.

The VERITAS observations of the bright flares from \lsi{} in 2014 provide constraints on the physical properties of the system around the acceleration region. The upper limit derived on the magnetic field strength following the prescription of \cite{2008MNRAS.383..467K} results in a Larmor radius that is a sizeable fraction of the total system size. This implies that particle leakage from the system will be non-negligible, a problem that is not accounted for in the model. In addition, the hard electron spectral index required to produce the observed gamma-ray spectral indices results in some tension with the photon indices in the X-ray regime. Both of these issues suggest that it is difficult for this type of simple model to reproduce the emission observed from the source. A beamed model could more easily account for the emission, hinting in favor of a $\mu$Q scenario for \lsi{}, although a jet-like structure could also be produced in PB systems \citep[e.g.,][]{2008MNRAS.387...63B}. A $\mu$Q scenario is also mildly favored by calculations of the mass of the compact object that place the minimum mass at about $2.5M_\sun$ \citep[e.g.,][]{0004-637X-519-1-336,2011ApJ...731..105N}.

There are a number of insights that could aid in distinguishing between the $\mu$Q and PB scenarios. High-resolution radio imaging could reveal extended jet-like structures that would point to a $\mu$Q scenario. Refined measurements of the orbital parameters of the system constraining the inclination angle and hence the mass of the system could clearly identify the compact object as a black hole if its mass were found to be $>3M_\sun$. The detection of pulsed emission from the source at any wavelength would unambiguously identify the compact object as a pulsar, but it is also possible that the dense stellar environment of the source could hinder such a detection. A correlation between the electromagnetic emission and H$\alpha$ variations would indicate that the emission is related to changes in the circumstellar disk of the optical star. It is clear that further observations of \lsi{} are necessary across all wavelengths to fully understand the nature of the varying emission from this source. 
\vspace{2ex}

\small{
This research is supported by grants from the U.S. Department of Energy Office of Science, the U.S. National Science Foundation and the Smithsonian Institution, and by NSERC in Canada. We acknowledge the excellent work of the technical support staff at the Fred Lawrence Whipple Observatory and at the collaborating institutions in the construction and operation of the instrument. The VERITAS Collaboration is grateful to Trevor Weekes for his seminal contributions and leadership in the field of VHE gamma-ray astrophysics, which made this study possible. A.\ O'FdB acknowledges support through the Young Investigators Program of the Helmholtz Association. A.W. Smith acknowledges support from the Fermi Cycle 7 Guest Investigator Program, grant number NNH13ZDA001N.
}

\end{document}